\documentclass[runningheads,a4paper]{llncs}

\usepackage{amssymb}
\usepackage{graphicx}
\usepackage{unitsdef} 
\usepackage{float}
\usepackage[dvipsnames]{xcolor}

\usepackage{tabularx}

\newif\ifdraft
\drafttrue

\ifdraft

\def\Hc#1{\textcolor{blue}{\textit{\textsf{ \small [HB: #1]}}}}  
\def\Hd#1{\textcolor{purple}{\textit{[deleted: #1]}}}  

\else

\def\Hc#1{}  
\def\Hd#1{}  
\fi

\begin{document}

\mainmatter  

\title{U-Net with spatial pyramid pooling for drusen segmentation in optical coherence tomography}

\titlerunning{Pyramid U-Net for drusen segmentation}

\author{Rhona Asgari 
%
\thanks{fatemeh.asgari@meduniwien.ac.at}%
\and Sebastian~Waldstein
\and Ferdinand~Schlanitz
\and Magdalena~Baratsits
\and Ursula~Schmidt-Erfurth
\and Hrvoje~Bogunovi\'c
}

\authorrunning{R. Asgari et al.}

\institute{Christian Doppler Laboratory for Ophthalmic Image Analysis, Department of Ophthalmology, Medical University of Vienna,
Vienna, Austria}

\toctitle{}

\tocauthor{}
\maketitle

\begin{abstract}
The presence of drusen is the main hallmark of early/inter\-mediate age-related macular degeneration (AMD). Therefore, automated drusen segmentation is an important step in image-guided management of AMD. There are two common approaches to drusen segmentation. In the first, the drusen are segmented directly as a binary classification task. In the second approach, the surrounding retinal layers (outer boundary retinal pigment epithelium (OBRPE) and Bruch's membrane (BM)) are segmented and the remaining space between these two layers is extracted as drusen. In this work, we extend the standard U-Net architecture with spatial pyramid pooling components to introduce global feature context. We apply the model to the task of segmenting drusen together with BM and OBRPE. The proposed network was trained and evaluated on a longitudinal OCT dataset of 425 scans from 38 patients with early/intermediate AMD. This preliminary study showed that the proposed network consistently outperformed the standard U-net model.
\end{abstract}
\section{Introduction}
Age-related macular degeneration (AMD) is a devastating retinal disease and a leading cause of blindness in the elderly population in the developed world~\cite{Wong2014}. The clinical hallmark and usually the first finding of AMD is the presence of waste deposits, called drusen. In the early stages, these drusen begin to accumulate in between two anatomical layers of the retina, the outer boundary retinal pigment epithelium (OBRPE) and the Bruch's membrane (BM). The drusen buildup and the consequent AMD progression to late stages are remarkably variable among affected individuals, resulting in its management being one of the biggest dilemmas in ophthalmology~\cite{Schlanitz2017}. Currently, the patient scheduling frequency is primarily guided by the amount of drusen, which is subjectively assessed by drusen segmentation in optical coherence tomography (OCT). OCT is the state-of-the-art imaging modality for assessing the retina in AMD. This fast and non-invasive acquisition technique allows to inspect the retina at a micrometer resolution, granting the possibility to study not only the retinal layers but also several disease-related abnormalities, including drusen. Manual drusen segmentation is very time consuming, which creates a need for advanced medical image computing methods that can measure distinct and pathognomonic changes in drusen morphology in an accurate, objective and reproducible manner.

\paragraph{\textbf{Related work}} 
In recent years, deep learning based and non deep learning based methods were applied on this task~\cite{GorgiZadeh2017,Khalid2017,novosel2017joint,fang2017automatic,shah2018multiple}. Generally it has been shown that deep learning based methods, namely convolutional neural networks (CNN), outperform the previous cost-function based models~\cite{GorgiZadeh2017,fang2017automatic,shah2018multiple}. In~\cite{GorgiZadeh2017} a basic U-Net is applied on drusen and layer segmentation. In~\cite{fang2017automatic} a combination of a CNN, graph search based methods and standard classifier is introduced. In~\cite{shah2018multiple} a retina layer segmentation task is tackled by a B-scan level CNN.

Drusen segmentation task can be tackled by segmenting the neighbouring layers in the retina: BM and OBRPE. An alternative approach is to segment drusen as an additional class. Our assumption is that this additional class will not only provide more information about the layers adjacent to drusen class, but will also help the network to characterize the appearance of both drusen and non-pathological regions where OBRPE and BM overlap. The size of drusen varies, meaning a given drusen could either be a small drusen at an early stage or a large drusen at a later stage. This point is not taken into account by a normal CNN applied on drusen segmentation. This can cause the network to miss drusen that are particularly small or, conversely, drusen that exceed the network's receptive field (figure~\ref{fig:samples}). In addition, retinal layers strictly follow the same topological ordering and drusen has to appear strictly in-between OBRPE and BM. 
In CNN models, contextual information and the spatial relation between different anatomical parts of the retina might be overlooked by the small receptive field of a CNN. The limitations of receptive fields in a CNN is discussed in more details in~\cite{Zhao2016,Kaiming2017}.

\begin{figure}
\centering
\includegraphics[width=0.9\textwidth]{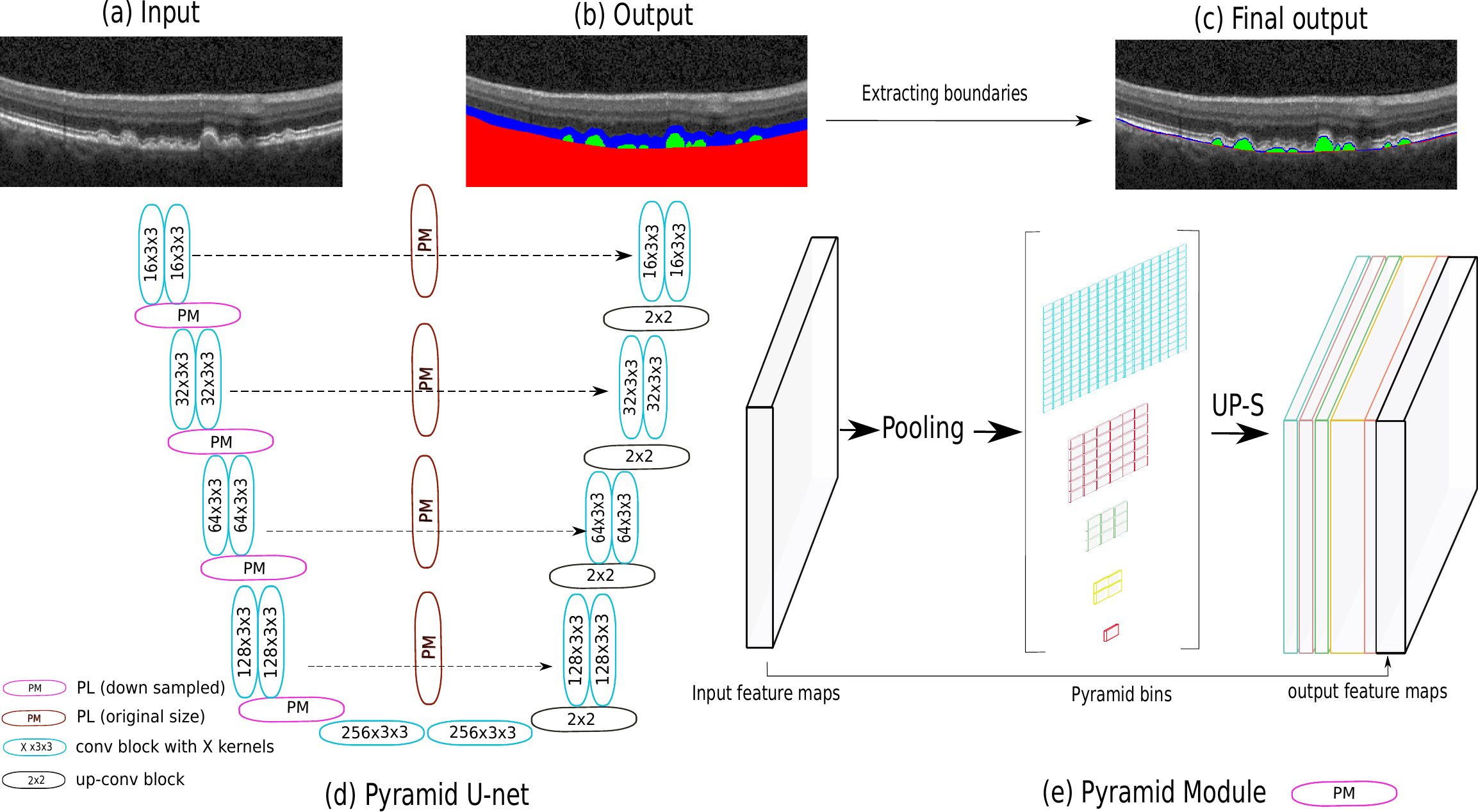}
\caption{(a) An input B-scan. (b) Corresponding network prediction output. (c) Final output after the post-processing. (d) The proposed model architecture. PM shows a pyramid module that is applied on the feature maps before they are passed to the next level. In the PM of each convolutional block on the decoder (pink PM), the reference size for output feature maps is half that of the input image size. In fact, the size of the feature maps in each layer of the decoder is halved. In the PM used for skip connections (Brown PM), the reference size is the same as the that of the input image. Therefor, the size of the feature maps passing through the skip connections does not change. (e) Five level pyramid module. The feature maps are gone through PM in order to have five-level bins. These five-level feature maps after up-sampling (UP-S) are concatenated with the original feature maps. }
\label{fig:pyramidUnet}
\end{figure}

A solution is to increase the receptive field in the CNN architecture. This could be approached in different ways, e.g. by a dilated convolution~\cite{yu2015multi}). In Pyramid scene parsing network (PSPNet), this is solved by a pyramid pooling module~\cite{Zhao2016}. Pyramid pooling is applying pooling with different window sizes. The idea is instead of having one size pooling with common kernel size of $2 \times 2$ resulting in halved size feature maps, applying pyramid pooling layer with different kernel size resulting in a sets of bins in a pyramid order (for example $1 \times 1$, $2 \times 2$, $3 \times 3$, $6 \times 6$). The coarsest pyramid level~(\(1 \times 1\)) resembles global pooling that covers the entire image (see Fig.~\ref{fig:pyramidUnet}(e)). Spatial pyramid pooling is also used in~\cite{Kaiming2017,gu2018deepdisc}.

In~\cite{Zhao2017,Kaiming2017,gu2018deepdisc} a spatial pyramid pooling layer is used once at the end of the last convolutional layer of the network. In this paper we take one step further and use a spatial pyramid pooling layer after each convolutional block of the encoder of a standard U-Net. We also evaluate the result of segmenting three classes instead of two classes, i.e. considering drusen as an additional, extra class. Finally, we use a weighted loss function to train our proposed model. We evaluate the performance of our approaches on the task of drusen and layer segmentation in retinas imaged with OCT. Results showed that the introduced model outperforms the baselines in term of Dice index of drusen segmentation, while also producing accurate delineations of the BM and the OBRPE surfaces.

\section{Methods}
U-Net~\cite{Ronneberger2015} has proven to be a suitable architecture for medical images, as it uses skip-connections to pass the feature maps from the encoder at the same level during the reconstruction stage, which makes the model convenient for segmentation tasks where precise location is needed. 
Thus, we chose U-Net as a backbone for our proposed pyramid U-Net with input image size of $256 \times 256$. 

A retina OCT scan is comprised of sequential 2D B-scans. Usually, segmentation algorithms detect the drusen boundaries in B-scans by segmenting the outer RPE and BM surfaces, as opposed to segmenting the drusen directly. In order to provide more information to the network, in this work we define a four-class segmentation task: \textit{Drusen}, \textit{RPE region}, \textit{BM region} and \textit{Background} (figure~\ref{fig:pyramidUnet} (b)). This is our first implemented approach and we evaluate whether adding the extra class helps the network to learn how the drusen class interacts with the neighboring classes.

In case of unbalanced classes, it is crucial to have a weighted loss function when evaluating multi-class segmentation output. In our work, drusen class pixels represent a very small fraction of the total pixels in an image. Thus, in order to handle the class imbalance, we use the following loss function to train a network that is based on Generalized Dice Coefficient~\cite{Crum2006} :

\begin{equation}\label{weight_equation }
  -2\frac{\Sigma_{c=1}^{4} \; \omega_{c} \; \Sigma_{n} y_{pred_{cn}} y_{true_{cn}} }{\Sigma_{c=1}^{4} \; \omega_{c} \; \Sigma_{n} y_{pred_{cn}} +  y_{true_{cn}}} \; .
\end{equation}

where \(y_{pred_{cn}}\) is the prediction by the network and \(y_{true_{cn}}\) is the ground truth image. $c$ is the number of classes, which in our proposed case is 4 (drusen, BM , OBRPE, and the background). \(\omega_{c}\) shows the weight attributed to a class \(c\) which is usually the inverse of the contribution of class \(c\) in data space. For the examined dataset,  \(\omega_{c}\) is set to 70, 20 and 10 for drusen class, OBRPE class and BM class respectively.

\subsection{Pyramid Module} \label{pyramidlayer}
Fig.~\ref{fig:pyramidUnet} shows the architecture of our proposed model. Each convolutional block is composed of two convolutional layers with $3\times3$ convolutions. Each convolutional block in the encoder is followed by one Pyramid Module (PM). A PM is composed of 5 different pooling levels with bins of size (\(1 \times 1\)), (\(2 \times 2\)), (\(3 \times 3\)),(\(6 \times 6\)) and (\(16 \times 16\)). The five-level pyramid module forms five separate sets of feature maps, each with a different size. Thus, in the first level of the network there are 5 sets of \(32 \times 256 \times 256\) feature maps (Fig.~\ref{fig:pyramidUnet}(e)), i.e., one series of feature maps for each pooling size. We apply average pooling with kernel size $pool\_size$ on these feature maps in each pyramid level in order to have results with bin size $(1 \times 1), (2 \times 2), (3 \times 3), (6 \times 6), (16 \times 16)$, respectively. 

In each level, a series of feature maps is followed by a separate \(1 \times 1\) convolutional to reduce the dimensionality of the feature maps to $n$. In this paper, $n$ is set to $32/2=16$ for the bin (\(2 \times 2\) bin) and to $32/4=8$ in the remaining pyramid levels (\(1 \times 1\), \(3 \times 3\), \(6 \times 6\) and \(16 \times 16\)). In each pyramid level, pooling kernel size will be calculated as: $pool\_size  = (input\_shape / bin\_size)$ in order to get feature maps with the target bin size. Since we are using U-Net as a baseline, where encoder uses \(2 \times 2\) max pooling in each level of the network, we keep the feature maps at each level of the network the same size as those in the basic U-Net. Therefore, all the feature maps of the different bin sizes are combined with the feature maps obtained by the pooling with kernel size \(2 \times 2\). The idea is that the feature maps from different bin sizes will add additional global context information to the main \(2 \times 2\) pyramid. The same rule applies for the following levels.  If $N$ is the number of the feature maps in each level and $n$ the desired number of the feature maps from a pooling bin $b \times b$, $n$ is set to $N/2$ for pooling with size \(2 \times 2\) and to $N/4$ for the rest of the pooling bins in the pyramid. 

\begin{figure}
\centering
\includegraphics[height=4.2cm, width=0.8\textwidth]{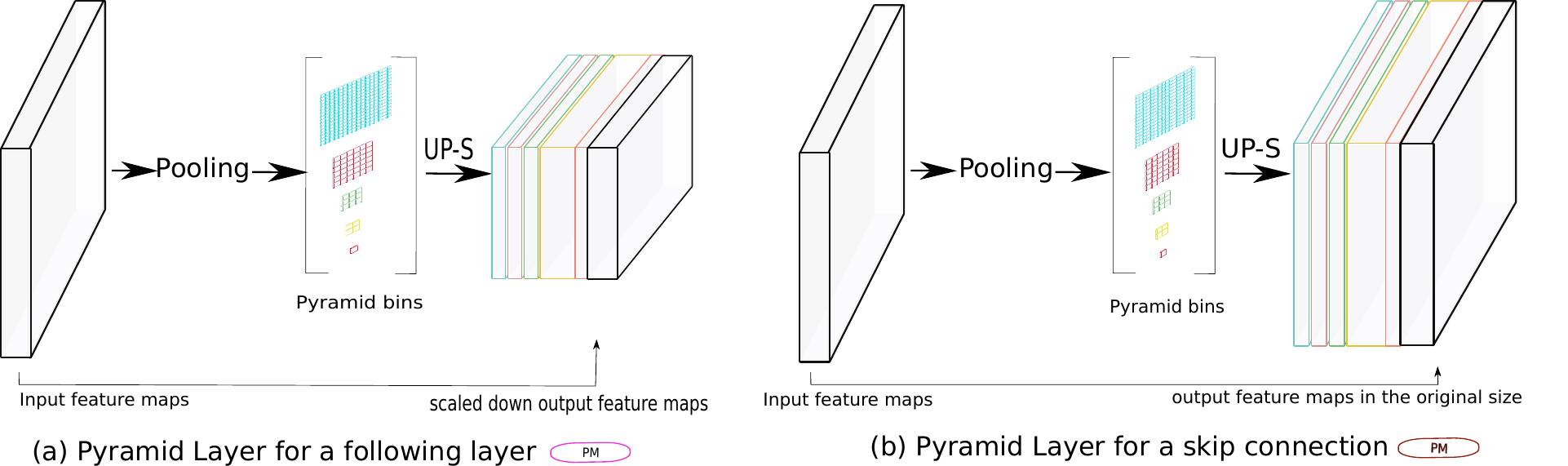}
\caption{(a) Pyramid module with the reference size of half of the input image size (b) Pyramid module with the reference size of the input image size.}
\label{fig:pyramidlayers}
\end{figure}

After applying the $1 \times 1$ convolution in each pooling level $p \times p$, there are 5 sets of feature maps of different sizes. In order to be able to concatenate these feature maps, each series of feature maps is up-sampled to the reference size of $ S$. For the pyramid module in the decoder, $ S$ is set to $1/2 \times(input\_size)$ . The feature maps at each level of the decoder are concatenated with the feature maps resulting from (\(2 \times 2\)) max pooling~(Fig.~\ref{fig:pyramidlayers}(a)). After concatenation, these pyramid feature maps are concurrently fed into the next layer. Conversely, for the pyramid modules on the skip connections $ S$ is set to $1\times(input\_size)$. Therefore, these feature maps keep their original dimension (Fig.~\ref{fig:pyramidlayers}(b)). The output of a PM (original feature maps and feature maps from 5 level bins) are simultaneously passed through the skip connections to the matching layer in the decoder.

The generalized Dice loss function was utilized for training the network. The predicted labels were regions for each target class (figure~\ref{fig:pyramidUnet} (b)). To acquire the final surfaces of the BM/OBRPE layers, a postprocessing strategy was applied. In each vertical column in the B-scan (called A-scan), the first row of activated pixels was extracted from the predicted BM region as the BM surface boundary. Similarly, in each vertical column in the B-scan (called A-scan), the last row of activated pixels was extracted from predicted OBRPE region as the OBRPE surface boundary.

\section{Experimental setup}
 \paragraph{Dataset} To train and evaluate the networks we use a private OCT dataset containing 425 OCT scans from 38 patients. We split the data into 34000 B-scans for training and validation (31 patients) and 7000 B-scans for testing (7 patients). Scans from the same subjects were always placed in the same set. Scans were acquired with Spectralis (Heidelberg Engineering, Heidelberg, Germany), which acquires anisotropic images with $1024\times97\times496$ voxels, each with the size of $5.7\times60.5\times3.87$ \micro m$^3$, and covering the field of view of $6\times6\times2$ mm$^3$.
 
\paragraph{Reference Standard}
 Each B-scan of every volume has been manually annotated in the following way. The Iowa Reference Algorithm~\cite{chen2012three} was first applied to generate a layer segmentation. The output was then manually corrected by an expert optometrist. Then, BM, OBRPE and the drusen regions are extracted from these annotations and used for training the network (Fig.~\ref{fig:pyramidUnet}). 
\paragraph{Training Setup} Our method and the baselines~\cite{GorgiZadeh2017,Zhao2016} were trained with a batch size of 16 iterated for 50 times, using Adam optimization with an initial learning rate of $\eta = 10^{-5}$. Input B-scans are normalized to zero mean and unit variance and resized to 256$\times$256 pixels. Based on equation~\ref{weight_equation }, \(\omega_{c}\) is set at 70, 20 and 10 respectively for drusen, \textit{RPE region} and \textit{BM region} in both baseline models and the introduced architecture. 

\begin{figure}[h!]
\centering
\includegraphics[width=0.8\textwidth]{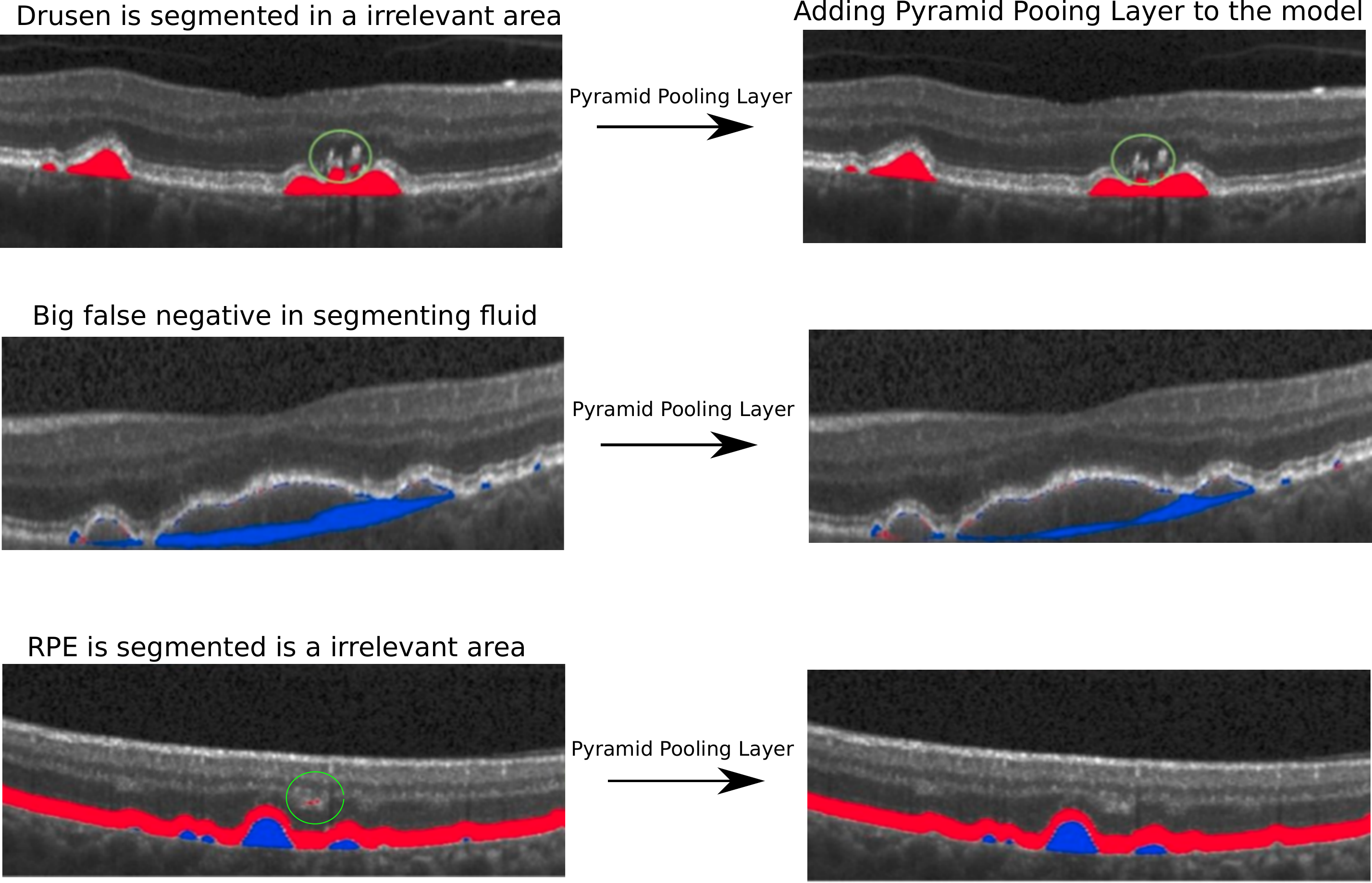}
\caption{(Left) Output of the basic U-Net~\cite{GorgiZadeh2017} and (Right) output of the proposed model. First row: The basic U-Net has erroneously segmented the drusen as the bright areas above RPE, although there should be no drusen on top of the RPE. Second row (blue is false-negative drusen): in the basic U-Net's output, drusen region exceeds the network's receptive field. Third row (drusen in blue and RPE in red): The part of the image which is completely outside of the outer retina is segmented as RPE by the basic U-Net.}
\label{fig:samples}
\end{figure}

\section{Results}
In order to evaluate our model, we compare it to several baselines. The first baseline is the standard U-Net architecture with two classes, BM and OBRPE, which has also been applied in the task of drusen segmentation by~\cite{GorgiZadeh2017}. We denote this baseline as UNet-2C in Table~\ref{table:dicetable}. The second baseline is the U-Net with drusen introduced as an extra class. In this baseline, instead of extracting the area between BM and OBRPE as drusen, drusen is specifically segmented as an extra class. We denote this baseline as UNet-3C. Finally, our proposed model has in addition a spatial pyramid pooling layer at each level of the basic U-Net, and is denoted as UNet-PPM (Table~\ref{table:dicetable}).

\begin{table}[h!]
  \begin{center}
    \caption{Quantitative evaluation. Patient-level mean Dice coefficient for drusen region segmentation and mean absolute error (MAE) in pixels for BM and OBRPE surface segmentation.} 
    \label{table:dicetable}
    \begin{tabular}{c|c|c|c}
    
      \textbf{Method} & \textbf{Dice}(Drusen)  & \textbf{MAE}(RPE) & \textbf{MAE}(BM)   \\
      
      \hline
      UNet-2C~\cite{GorgiZadeh2017} & 70.25  & 1.42 & 1.35 \\
      \hline
      UNet-3C& 72.20 & 1.27 & 1.21 \\
      \hline
      UNet-PPM &  \textbf{74.73} & \textbf{0.79} & \textbf{0.71} \\
      
    \end{tabular}
  \end{center}
\end{table}

\begin{figure}
\centering
\includegraphics[width=0.8\textwidth]{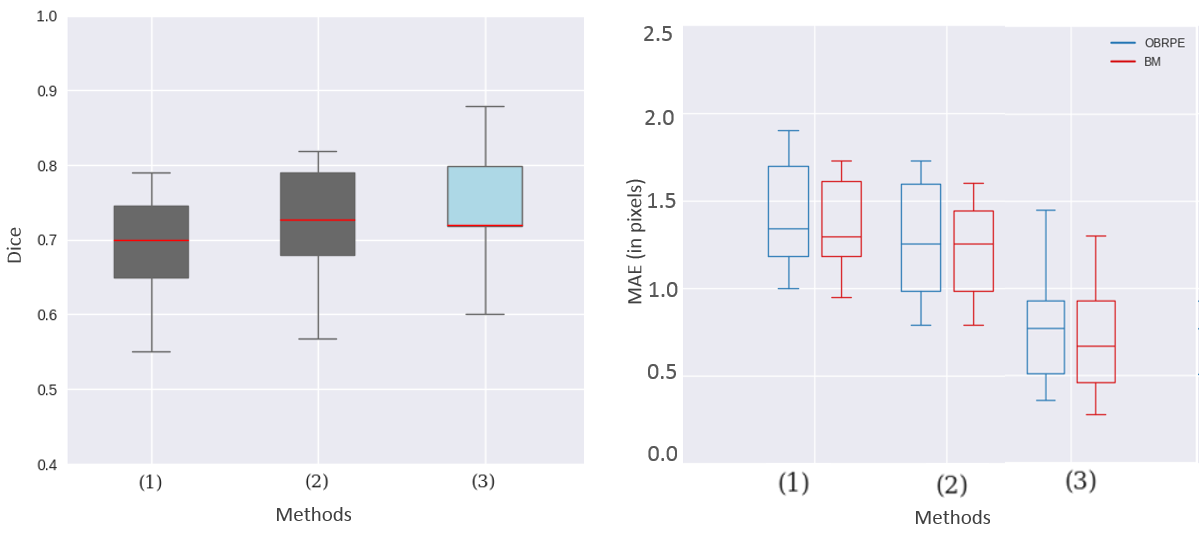}
\caption{
Segmentation performance of different models on drusen (left), and OBRPE and BM (right). (1) UNet-2C~\cite{GorgiZadeh2017}: U-Net with two classes, BM and OBRPE. (2) UNet-3C: original U-Net with three classes, BM, OBRPE and drusen. (3) UNet-PPL: the proposed model with three classes and pyramid pooling layers.}
\label{fig:boxplot}
\end{figure}

An example of segmentation output is shown in Fig.~\ref{fig:samples}. It shows how the pyramid pooling method solves some fundamental issues in drusen segmentation by adding global contextual information to the feature maps which are being transferred through the network. We quantitatively evaluated the segmentation performance of the drusen, OBRPE and BM segmentation. Table~\ref{table:dicetable} shows the results of this evaluation, per patient dice coefficient for drusen segmentation and mean absolute error for OBRPE and BM. In addition, figure~\ref{fig:boxplot} shows a box-plot of per patient dice coefficient for drusen and mean absolute error for BM and RPE segmentation. One can observe that by using the pyramid module, our proposed method was able to outperform the other baseline networks.




\section{Discussion}
Utilizing global spatial context is crucial for avoiding anatomically impossible segmentation such as finding drusen above RPE instead of below it. It is still a challenge to learn the plausible spatial relationships between object classes from a training dataset using statistical machine learning approaches. We proposed incorporating the pyramid pooling module into U-Net. The results showed that the proposed extension utilized the larger context for segmentation and clearly outperformed the baseline U-Net model. The proposed method is an important step towards the accurate quantification of drusen, crucial for the successful clinical management of patients with early AMD. Finally, given the widespread use of U-Net for medical image segmentation in general, the proposed extension would have an impact beyond its application in drusen segmentation.

\paragraph{}This work was funded by the Christian Doppler Research Association, the Austrian Federal Ministry for Digital and Economic Affairs and the National Foundation for Research, Technology and Development. We thank the NVIDIA corporation for a GPU donation.

%
%
\bibliographystyle{splncs}
\bibliography{bibs/IEEEabrv,bibs/_editors,bibs/_confAbrv,bibs/_journalsAbrv,bibs/main}

\end{document}